\documentstyle[12pt]{article}
\begin{document}
\baselineskip=15pt
\begin{center}
\bf {ENTROPY AND HADAMARD MATRICES}\\
~~\\
~~\\
H. Gopalkrishna Gadiyar$^{*}$, K M Sangeeta Maini$^{*}$, R. Padma$^{*}$ and \\
H.S. Sharatchandra$^{**}$
\end{center}

\begin{flushleft}
\baselineskip=15pt
$^{*}$ AU-KBC Research Centre,
M.I.T. Campus, Anna University,
Chromepet,\\
~~Chennai 600 044 INDIA\\
~~E-mail: gadiyar@au-kbc.org, sangeeta\_maini@au-kbc.org,\\
~~~~~~~~~~~~\,padma@au-kbc.org
~~\\
$^{**}$ Institute of Mathematical Sciences,
C.I.T. Campus, Taramani P.O.,\\
~~~Chennai 600 113 INDIA\\
~~~E-mail: sharat@imsc.ernet.in
\end{flushleft}

\baselineskip=15pt
\begin{center}
{\bf{Abstract}}
\end{center}

The entropy of an orthogonal matrix is defined. It provides a new interpretation of Hadamard matrices as those that saturate the bound for entropy. It appears to be a useful Morse function on the group manifold. It has sharp maxima and other saddle points. The matrices corresponding to the maxima for 3 and 5 dimensions are presented. They are integer matrices (upto a rescaling.)

\newpage

Hadamard's maximum determinant problem asks for $n \times n$ matrix $h_n$ with elements of magnitude $\le 1$ which has the maximum possible magnitude for the determinant. Hadamard's inequality is $\vert \mbox{det}(h_n) \vert \le n^{\frac{n}{2}}$. The equality is attained for some values of $n$, and the corresponding matrices are called Hadamard matrices $H_n$. They are orthogonal matrices upto a proportionality factor $\sqrt{n}$:
$$
H_n H_n^T = n I_n \, , \hspace{10cm} (1)
$$
and have elements $\pm 1$. Geometrical interpretation of the maximum determinant problem is to look for $n$ vectors from the origin contained within the cubes $\vert x_i \vert \le 1$, $i=1,..,n$ and forming a rectangular parallelopiped of maximum volume. In case of Hadamard matrices, these vectors are a set of main diagonals, which obviously have the maximum possible lengths. Moreover they form an orthogonal set to provide the maximum volume. In other dimensions, it is not possible to have mutually orthogonal main diagonals. Nevertheless, we expect that the maximum determinant is from vectors which are `close' to some main diagonals. This means that the entries of the matrix have magnitudes close to $1$. Also we might expect a continuous set of matrices which have the maximum determinant because certain allowed changes in the vectors would increase their lengths and decrease the angles between them keeping the volume unchanged. 

Hadamard matrices find natural applications in error-correcting codes [1] and provide ``biorthogonal" codes. There is extensive work on finding matrices with entries $\pm 1$ which have maximum determinant. This is a subclass of the Hadmard's problem. There are also many generalizations and relatives of the problem. For example, [2] studies the computationally difficult problem of finding a largest $j$-dimensional simplex in a given $d$-dimensional cube.

In this note we use entropy to give a new criterion for Hadamard matrix and obtain its generalization. We define entropy of orthogonal matrices and Hadamard matrices (appropriately normalized) saturate the bound for the maximum of the entropy. The maxima (and other saddle points) of the entropy function have intriguing structure and yield generalizations of Hadamard matrices. It appears that entropy is a very useful Morse function on the group manifold and not only the number [3] but also the location of the saddle points have interesting features.

Consider $n$ random variables with a set of possible outcomes $i=1,..,n$ having probabilities $p_i$, $i=1,..,n$. We have ${\displaystyle{\sum_{i=1}^n ~p_i~=~1}}$. The (Shannon) entropy is
$$
H\{ p_i \} = \sum_{i=1}^n p_i ~ \ln \frac{1}{p_i} \hspace{9cm} (2)
$$
This has the minimum value zero for the case of certainty,
$$
p_i=\left\{ \begin{array}{ll}
1 & i~=~j~\mbox{for some} ~j,\\
0 & i~\neq j\,. \end{array} \right.
$$
It has the maximum value $\ln n$ when all outcomes are equally likely, 
$$
p_i=\frac{1}{n} \, , ~\forall i=1,..,n.
$$

We now define entropy of an orthogonal matrix $O_{~j}^i$, $i, j~=~1,~2,...n$. Here $O_{~j}^i$'s are real numbers with the constraint
$$
\sum_{i=1}^n O^i_{~j}~O^i_k~=~\delta _{jk} \, . \hspace{9cm} (3)
$$
In particular, $i^{th}$ row of the matrix is a normalized vector for each $i=1,..,n$. We may associate probabilities $p^{(i)}_j~=~(O_{~j}^i)^2$ with the $i^{th}$ row, as ${\displaystyle{\sum_{j=1}^n ~p_j^{(i)}~=~1}}$, for each $i$. We define the (Shannon) entropy for the orthogonal matrix as the sum of the entropies of each row:
$$
H\{O_{~j}^i\} = -\sum_{i,j=1}^n (O^i_{~j})^2 ~ \ln (O^i_{~j})^2 \, .
$$
The minimum value zero is attained by the identity matrix $O_{~j}^i~=~\delta _j^i$ and related matrices obtained by interchanging rows or changing the signs of the elements. The entropy of the $i^{th}$ row can have the maximum value $\ln n$, which is attained when each element of the row is $\pm \frac{1}{\sqrt{n}}$. This gives the bound, $H\{O_{~j}^i\} \le n~\ln n$. In general, the entropy of an orthogonal matrix cannot attain this bound because of the orthogonality constraint (3) which constrains $p^{(i)}_j$ for different rows. In fact the bound is obtained only by the Hadamard matrices (rescaled by $n^{-\frac{1}{2}}$). Thus we have a new criterion for the Hadamard matrices (appropriately normalized): those orthogonal matrices which saturate the bound for entropy.

Note that the entropy is large when each element is as close to $\pm \frac{1}{\sqrt{n}}$ as possible, i.e., to a main diagonal. Thus maximum entropy condition is similar to the maximum determinant condition of the Hadamard. Moreover we find that the peaks of entropy are isolated and sharp in contrast to the determinant. Also, the matrices corresponding to the maxima have very interesting features even for those dimensions $n$ for which Hadamard matrices do not exist. We have obtained matrices maximizing the entropy for $n~=~3$ and $5$ by numerical computation. For $n~=~3$, the matrix is 
$$
\left[ \begin{array}{rrr}
-\frac{1}{3} & \frac{2}{3}& \frac{2}{3}\\
~&~&\\
\frac{2}{3} & -\frac{1}{3}& \frac{2}{3}\\
~&~&\\
\frac{2}{3} & \frac{2}{3}& -\frac{1}{3}
\end{array}
\right] \hspace{9cm} (4)
$$
The matrix entries are all rational numbers and with a rescaling of each row by $3$ we get an integer matrix. This was unexpected. For $n=5$, the result is similar: the magnitudes of the elements in each row are $\frac{2}{5}$ repeated $4$ times and a $\frac{3}{5}$. This set can be generalized for any $n$: 
$$
\frac{n-2}{n}, \underbrace{\frac{2}{n}, \cdots , \frac{2}{n}}_{(n-1) \mbox{times}} \,  \hspace{9.5cm} (5)
$$
A vector with these components has unit normalization as a consequence of the identity,
$$
n^2=(n-2)^2+2^2(n-1)
$$
We expect that for each $n \ge 2$, there exist orthogonal matrices with entries of magnitude $\frac{2}{n}$ repeated $(n-1)$ times along with one $\frac{n-2}{n}$ in each row. Only the location of the element $\frac{n-2}{n}$ and the signs of the elements change from row to row. Nonetheless it is easy to argue that this cannot be the maximum for all $n$. For $n$ very large the rows are close to one of the axes, and not to a main diagonal. Therefore the entropy is close to zero instead of being close to the bound $n \ln n$. We expect that different maxima emerge as $n$ increases. It will be interesting to find the matrix corresponding to the maximum entropy for large $n$, but it is numerically difficult. 

Indeed, for $n~=~4$, the Hadamard matrix and not the family (4) gives the maximum. Numerical computation has shown that even in this case, member of the family (5) is an extremum of the entropy, though not the maximum. 

The picture that emerges is as follows. For each $n$, there are saddle points apart from maxima and minima. For example, for $n~=~3$ there is a saddle point and the corresponding matrix is:
$$
\left[ \begin{array}{rrr}
\frac{1}{2} & \frac{1}{\sqrt{2}}& \frac{1}{2}\\
~&~&\\
\frac{1}{\sqrt{2}} & 0& -\frac{1}{\sqrt{2}}\\
~&~&\\
\frac{1}{2} & -\frac{1}{\sqrt{2}}& \frac{1}{2}
\end{array}
\right]  \hspace{8cm} (6)
$$
The entropy is peaked quite sharply at all extrema. As we change $n$, new maxima may emerge as for $n~=~4$. There will also be crossovers among the families of saddle points. 

Thus entropy function has a rich set of sharp extrema. It appears to be a useful Morse function for the $O(n)$ group manifold. (It is possible to define the corresponding object for other group manifolds too.) Even the location of the saddle points i.e., the structure of the corresponding matrix is interesting. To our knowledge entropy has not been applied as a Morse function in the past. (It has been applied for a classification of the Bernoulli shifts [4].)

The specific way in which the matrices (4), (6) solve the equations for the extrema is interesting. It turns out that they are simultaneously extrema of a class of entropy functions called Renyi entropy:
$$
H_a\{O_{~j}^i\}=\sum_{i,j}((O_{~j}^i)^2)^a
$$
where $a > 0$. We can recover the Shannon entropy from $O(\epsilon )$ terms when $a~=~1-\epsilon $. To obtain the equations for the extrema we use a Lagrange multiplier $\lambda _{ij}$ symmetric in $i$ and $j$ for the constraint (1). The equations are
$$
((O_{~j}^i)^2)^{a-1}~O_{~j}^i~=~ \frac{2}{a} \sum_k \lambda_{jk} O_k^i \, .
$$
Using the orthogonality relation (1) we may rewrite this as   
$$
\sum_i ((O_{~j}^i)^2)^{a-1}~O_{~j}^i~O_{~l}^i=\frac{2}{a} \lambda_{jl}\, .
$$
For $j=l$, this simply gives the value of $\lambda _{jj}$. The nontrivial equations are obtained by using $\lambda_{jl}~=~\lambda_{lj},~j~\neq ~l$:
$$
\sum_i (( O_{~j}^i)^2)^{a-1}~O_{~j}^i~O_{~l}^i=\sum_i ((O_{~l}^i)^2)^{a-1}~O_{~l}^i~O_{~j}^i \, , ~l\neq j \, .
$$
We demonstrate how these nonlinear equations are satisfied by the maximum (4) for $n=3$. Using the elements in the first and second rows, we have, 
$$
\left( \frac{1}{3}\right)^{2(a-1)}\left( -\frac{1}{3}\right)\left(\frac{2}{3}\right)
+ \left( \frac{2}{3}\right)^{2(a-1)}\left( \frac{2}{3}\right)\left( -\frac{1}{3}\right)
+ \left( \frac{2}{3}\right)^{2(a-1)}\left( \frac{2}{3}\right)\left( \frac{2}{3}\right)\\
$$
$$
=\left( \frac{2}{3}\right)^{2(a-1)}\left( \frac{2}{3}\right)\left( -\frac{1}{3}\right)
+\left( \frac{1}{3}\right)^{2(a-1)}\left( -\frac{1}{3}\right)\left( \frac{2}{3}\right)
+\left( \frac{2}{3}\right)^{2(a-1)}\left( \frac{2}{3}\right)\left( \frac{2}{3}\right) \hspace{1cm} (7)
$$
We see that either $O_{~j}^i=\pm O_{~l}^i$ so that the contribution to the left hand side and the right hand side in equation (7) are same; or else, $O_{~j}^i=\pm O_{~l}^{i'}$ for some $i$ and $i'$ so that again the contribution to the left hand side and the right hand side are the same.

In this note we defined entropy of orthogonal matrices. We observed that it provides a new characterization of Hadamard matrices as those that saturate the bound for entropy. In dimensions where Hadamard matrices do not exist, entropy has sharp maxima (and other saddle points) in contrast to the determinant considered by Hadamard. We obtained matrices that maximze entropy in $3$ and $5$ dimensions. Surprisingly, they have rational entries. We argued that as $n$ is varied there are families of maxima and saddle points which have interesting cross overs. It is challenging to find the maxima for large $n$ (for non-Hadamard dimensions.) We expect it will be close to Hadamard matrices in structure.

\newpage
\noindent {\bf References}
\begin{enumerate}
\item[1] Mc Williams, F.J. and Sloane, NJA. The Theory of Error-Correcting Codes, New York, Elsevier; 1978.
\item[2] P. Gritzmann, V. Klee and D. Larman. Largest j-simplices in n-polytopes, Disc. Comp. Geom. {\bf 13}, 3/4, 1995, 477-515.\\
\item[3] R. Bott, J. Mather, Topics in topology and differential geometry, 460-515, Battelle Rencontres, 1967, Lecture in Mathematics and Physics, Ed. C.M. DeWitt and J.A. Wheeler, W.A. Benjamin, Inc. New York 1968.
\item[4] D.S. Ornstein, Bernoulli shifts with same entropy are isomorphic, Advances in Math. {\bf 4}(1970),337-352. 
\end{enumerate}  
\end{document}